\documentclass[aps,prapplied,reprint,superscriptaddress]{revtex4-2}

\usepackage{graphicx}
\usepackage{overpic}
\usepackage{amsmath,amssymb,bm}
\usepackage[colorlinks=true,linkcolor=blue,citecolor=blue,urlcolor=blue]{hyperref}
\usepackage[caption=false]{subfig}
\usepackage[normalem]{ulem} 

\begin{document}


\title{Flux-tunable transmon incorporating a van der Waals superconductor via an Al/AlO$_x$/4Hb-TaS$_2$ Josephson junction}

\author{Eliya Blumenthal}
\thanks{These authors contributed equally to this work.}
\affiliation{Department of Physics, Technion -- Israel Institute of Technology, Haifa 32000, Israel}
\affiliation{Helen Diller Quantum Center (Technion)}

\author{Ilay Mangel}
\thanks{These authors contributed equally to this work.}
\affiliation{Department of Physics, Technion -- Israel Institute of Technology, Haifa 32000, Israel}
\affiliation{Helen Diller Quantum Center (Technion)}

\author{Amit Kanigel}
\affiliation{Department of Physics, Technion -- Israel Institute of Technology, Haifa 32000, Israel}
\affiliation{Helen Diller Quantum Center (Technion)}

\author{Shay Hacohen-Gourgy}
\affiliation{Department of Physics, Technion -- Israel Institute of Technology, Haifa 32000, Israel}
\affiliation{Helen Diller Quantum Center (Technion)}

\begin{abstract}
Incorporating van der Waals (vdW) superconductors into Josephson elements extends circuit-QED beyond conventional Al/AlO$_x$/Al tunnel junctions and enables microwave probes of unconventional condensates and subgap excitations. In this work, we realize a flux-tunable transmon whose nonlinear inductive element is an Al/AlO$_x$/4Hb-TaS$_2$ Josephson junction. The tunnel barrier is formed by sequential deposition and full in-situ oxidation of ultrathin Al layers on an exfoliated 4Hb-TaS$_2$ flake, followed by deposition of a top Al electrode, yielding a robust, repeatable hybrid junction process compatible with standard transmon fabrication. Embedding the device in a three-dimensional copper cavity, we observe a SQUID-like flux-dependent spectrum that is quantitatively reproduced by a standard dressed transmon--cavity Hamiltonian, from which we extract parameters in the transmon regime. Across measured devices we obtain sub-microsecond energy relaxation ($T_1$ from $0.08$ to $0.69~\mu$s), while Ramsey measurements indicate dephasing faster than our $16$ ns time resolution. We also find a pronounced discrepancy between the Josephson energy inferred from spectroscopy and that expected from the Ambegaokar--Baratoff relation using room-temperature junction resistances, pointing to nontrivial junction physics in the hybrid Al/AlO$_x$/4Hb-TaS$_2$ system. Although we do not resolve material-specific subgap modes in the present geometry, this work establishes a practical route to integrating 4Hb-TaS$_2$ into coherent quantum circuits and provides a baseline for future edge-sensitive designs aimed at enhancing coupling to boundary and subgap degrees of freedom in vdW superconductors.
\end{abstract}

\maketitle

\section{Introduction}

Superconducting transmon qubits provide a mature circuit-QED (cQED) platform for coherent control and precision spectroscopy of mesoscopic superconducting phenomena~\cite{Koch2007Transmon}.
In most implementations, the Josephson element is an Al/AlO$_x$/Al tunnel junction, which offers reliable fabrication and low loss but restricts the accessible superconducting order parameters and quasiparticle spectra to those of conventional $s$-wave aluminum.

Van der Waals (vdW) quantum materials offer an avenue to extend cQED to correlated and topological superconductors, where the order parameter, low-energy quasiparticles, and boundary modes can differ qualitatively from conventional tunnel junctions. Recent work has demonstrated compact transmon implementations leveraging vdW heterostructures, including crystalline parallel-plate capacitors and fully vdW Josephson elements using a flake transfer method~\cite{Antony2021MiniaturizingTransmon, Balgley2025CoherentCompactVdWTransmons}. A practical requirement for bringing these materials into cQED is a repeatable junction process that forms a robust tunnel barrier on an exfoliated vdW surface and is compatible with standard transmon fabrication. As a concrete demonstration, we focus on the 4Hb polytype of TaS$_2$, which is a transition-metal dichalcogenide that forms a naturally occurring heterostructure of alternating 1T and 1H layers. It superconducts with $T_c \approx 2.7~\mathrm{K}$ and has been reported to exhibit signatures consistent with time-reversal symmetry breaking at $T_c$~\cite{Ribak2020Chiral4HbTaS2}.
Scanning tunneling spectroscopy has further reported zero-bias states at step edges and within vortex cores, consistent with boundary and subgap states in a topological nodal superconductor~\cite{Nayak2021Evidence}.
Recent phase-sensitive Little--Parks measurements in 4Hb-TaS$_2$ rings report $\pi$-shifted oscillations in a subset of devices, supporting non-$s$-wave pairing components~\cite{Almoalem2024_pi_shifts}.
At the same time, the low-energy gap structure remains under active discussion, including proposals of robust gapless superconductivity associated with multi-band pair breaking~\cite{Dentelski2021Robust} and recent microwave cavity studies of phase stiffness in few-layer devices~\cite{chistolini2025contactlesscavitysensingsuperfluid}.

These developments motivate the integration of 4Hb-TaS$_2$ into coherent superconducting circuits.
A transmon whose Josephson element interfaces Al and 4Hb-TaS$_2$ provides a controllable microwave probe of the condensate and of potential subgap degrees of freedom, complementary to tunneling microscopy and transport.
Here we take a first step toward this goal by realizing a flux-tunable transmon in which the Josephson element is an Al/AlO$_x$/4Hb-TaS$_2$ junction.
We demonstrate standard spectroscopy and time-domain control in a 3D cQED architecture, establishing an experimental baseline and a fabrication route for future edge-sensitive device geometries.

\section{Device design and fabrication}

Devices were fabricated on Si/SiO$_2$ substrates (250~$\mu$m Si, 285~nm thermal oxide) pre-patterned with an alignment grid of either Pt/Cr (20~nm/5~nm) or Nb (25~nm).
Substrates were solvent-cleaned by sequential sonication (5~min each in acetone, ethanol, and IPA) and baked at 180~$^\circ$C for 2~min.

Flakes of 4Hb-TaS$_2$ were exfoliated via a dry-transfer method onto the prepared substrates.
Candidate flakes were identified by optical microscopy and registered to nearby alignment markers.
The substrate was coated with a PMMA trilayer resist stack (two layers of 495 A4 followed by 950 A2), and patterned using electron-beam lithography (RAITH e-Line 150, 10~kV). It was then developed in 1:3 MIBK:IPA (90~s), and rinsed in IPA.
A short oxygen-plasma ashing step (30~s) was used to reduce resist residues in exposed areas.

Metal deposition was performed in an e-beam evaporator using a multi-step process designed to form a controlled tunnel barrier on exfoliated vdW crystals.
First, an in-situ Ar ion mill removed surface contamination and milled a few nanometers into the 4Hb-TaS$_2$ at $\sim$45$^\circ$ incidence to promote robust side-contact at junction interfaces.
A short Ti gettering evaporation was used to reach base pressure.
A nominal $\sim$1~nm AlO$_x$ barrier was produced by sequential deposition of 2--3 thin Al layers at $10^\circ$ tilt ($\sim5~\AA$ each at $\sim0.5~\AA$/s), each fully oxidized in situ (80~Torr, 5\% O$_2$ in Ar, 40~min) before the next deposition.
A top Al layer ($\sim$200~nm) was then deposited at the same tilt angle.
Lift-off was performed in NMP at 80~$^\circ$C for 3~h, followed by IPA rinse and N$_2$ dry.

\begin{figure}
    \centering
    \subfloat{\begin{overpic}[width=0.99\linewidth]{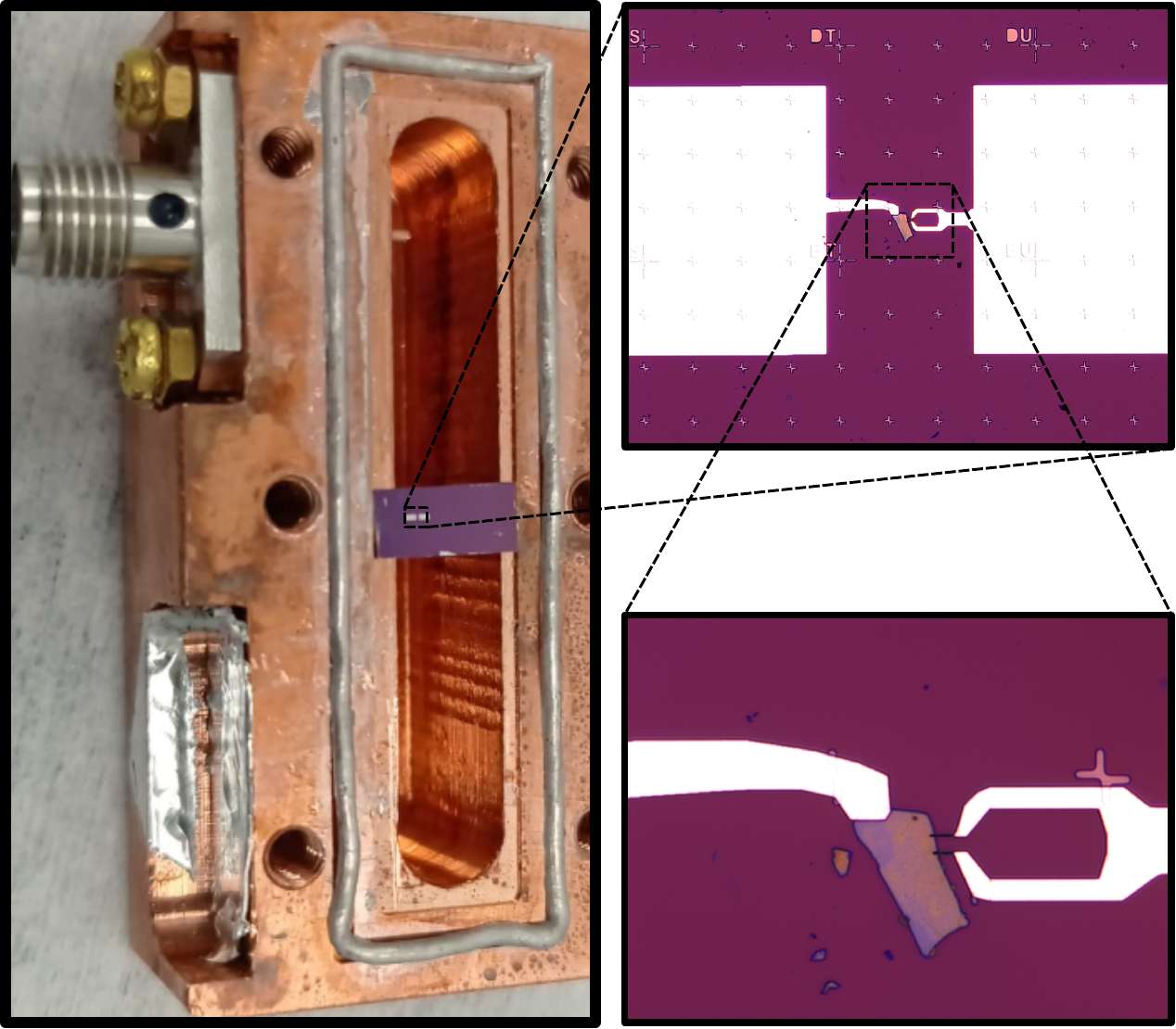}
    \put(2, 82){(a)}
    \end{overpic}}\\[5mm]
    \subfloat{\begin{overpic}[width=0.45\linewidth]{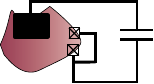}
    \put(2,55){(b)}
    \end{overpic}}
    \hspace{5mm}
    \subfloat{\begin{overpic}[width=0.45\linewidth]{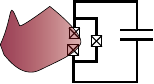}
    \put(2,55){(c)}
    \end{overpic}}
    
    \caption{(a) Photograph of the 3D copper cavity package and optical micrographs of a device 1B.
    (b,c) Schematic layouts of SQUID-transmon design 1 and 2. Design 1 routes supercurrent through the flake between the capacitor pads; Design 2 uses an asymmetric SQUID to reduce sensitivity to bulk transport through the flake.}
    
    \label{fig:fab}
\end{figure}

We implemented two SQUID-based transmon layouts (Fig.~\ref{fig:fab}).
In \emph{Design 1}, the SQUID loop is placed between one capacitor pad and the 4Hb-TaS$_2$ flake, while the second pad connects to the opposite side of the flake via a larger contact.
In \emph{Design 2}, an asymmetric SQUID geometry was used to reduce sensitivity to supercurrent pathways through the flake bulk: one branch contains hybrid Al/AlO$_x$/4Hb-TaS$_2$ junction element(s), while the other branch contains a conventional Al/AlO$_x$/Al junction element.
We fabricated and cooled down 2 samples of Design 1 and 3 samples of Design 2.

Individual devices were cleaved and mounted in a 3D OFHC copper cavity.
The cavity--qubit assembly was cooled to a base temperature of $\sim$10~mK in a dilution refrigerator.
Standard filtered and attenuated microwave lines provided cavity readout and qubit control.
A superconducting coil mounted above the cavity applied a DC magnetic field threading the SQUID loop to tune the effective Josephson energy.

\section{Results}
\begin{figure*}
    \centering
    \includegraphics[width=1\linewidth]{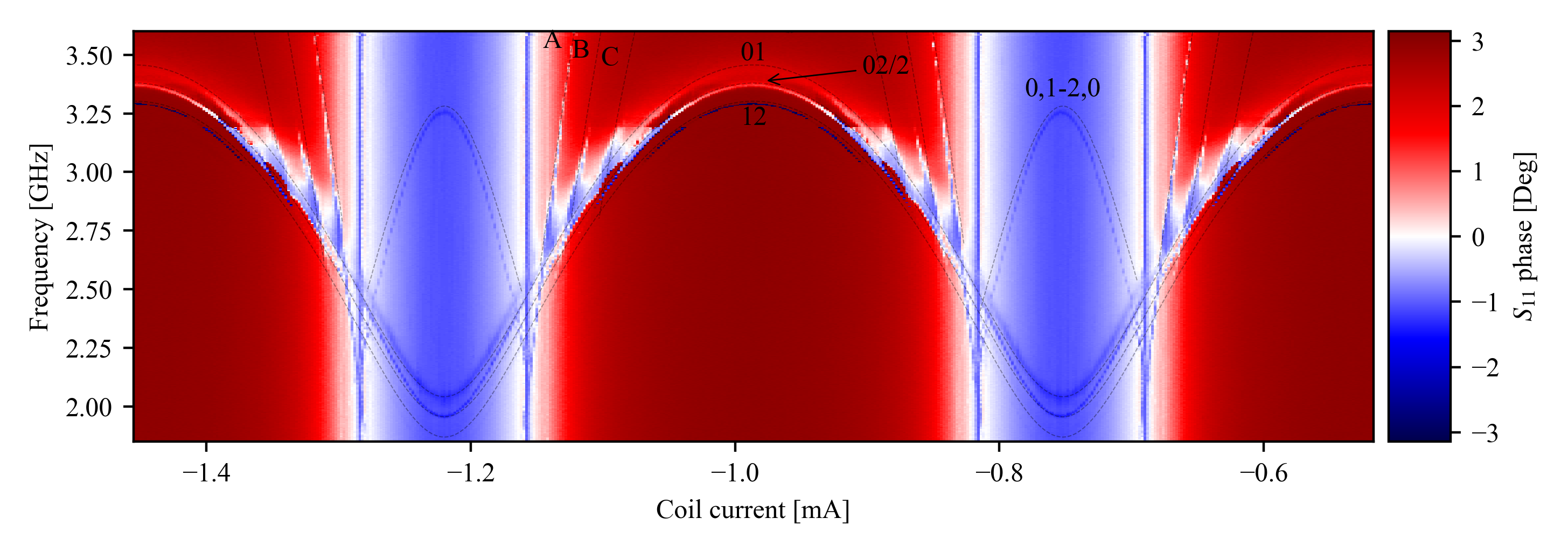}
    \caption{
    Two-tone spectroscopy as a function of coil current (external flux).
    Dashed curves: transition frequencies obtained from diagonalizing the dressed Hamiltonian in Eq.~\ref{eq:H} using extracted parameters.
    Weak additional features at high drive are consistent with cavity-assisted and multi-photon transitions within the dressed spectrum. The lines marked by A, B and C represent a Raman transition at a frequency of $(f_{1,0}-f_{0,0})+(f_{3,0}-f_{0,1})$, $(f_{1,0}-f_{0,0})+(f_{4,0}-f_{1,1})$ and $(f_{1,0}-f_{0,0})+(f_{5,0}-f_{2,1})$, respectively. The 01 transition of the transmon is power-broadened.
    }
    \label{fig:spectroscopy}
\end{figure*}
\subsection{Two-tone spectroscopy and parameter extraction}

We performed two-tone spectroscopy by driving the cavity near its fundamental mode while sweeping a second tone to probe transmon transitions.
Figure~\ref{fig:spectroscopy} shows a representative flux-dependent spectrum.
The observed transitions are accurately captured by a standard transmon--cavity Hamiltonian,
\begin{equation}
\hat{H} = \hbar\omega_c \hat{a}^\dagger \hat{a} + 4E_C \hat{n}^2 - E_J(\Phi)\cos\hat{\phi}
+ \hbar g (\hat{a}^\dagger + \hat{a})\hat{n},
\label{eq:H}
\end{equation}
where $\omega_c$ is the cavity frequency, $E_C$ the charging energy, $E_J(\Phi)$ the flux-dependent Josephson energy of the SQUID, $\hat{n}$ and $\hat{\phi}$ are conjugate charge and phase operators, and $g$ is the capacitive coupling rate.
For a SQUID, the Josephson energy is $E_J(\Phi) = E_{J,\Sigma}\cos(\pi\Phi/\Phi_0)\sqrt{1+d^2\tan^2\left(\pi\frac{\Phi}{\Phi_0}\right)}$, where $E_{J,\Sigma} = E_{J1}+E_{J2}$, $d=\left(E_{J2}-E_{J1}\right)/\left(E_{J1}+E_{J2}\right)$ and $\Phi_0$ is the magnetic flux quantum.

For the device 1A in Fig.~\ref{fig:spectroscopy}, we extract $E_C/h \approx 140~\mathrm{MHz}$,  $E_{J,\Sigma}/h \approx 11.6~\mathrm{GHz}$ and $d\approx 0.35$ corresponding to a maximum qubit frequency $\omega_{01}/2\pi \approx 3.46~\mathrm{GHz}$ and placing the device in the transmon regime ($E_J/E_C \gg 1$).

In addition to the primary transition, weaker lines appear at higher drive; these are consistent with multi-photon and cavity-assisted processes in the dressed transmon--cavity spectrum. 

Notably, the Ambegaokar--Baratoff relation~\cite{AmbegaokarBaratoff1964}
\begin{equation}
E_J=\frac{\Phi_0\Delta}{4R_n},
\end{equation}
with $\Delta$ the superconducting gap and $R_n$ the junction's normal-state resistance, is not satisfied in our devices that use only Al/AlO$_x$/4Hb-TaS$_2$ junctions. For device 1B we extract a total Josephson energy of $E_{J,\Sigma}/h \approx 13.7~\mathrm{GHz}$, while the junction resistances measured at room temperature are $2.4~\mathrm{k\Omega}$ (device 1A) and $1.9~\mathrm{k\Omega}$ (device 1B). Plugging these values into the relation yields an inferred gap of only $33$--$35~\mathrm{\mu V}$, which is inconsistent with the known gap of aluminum ($\sim 162~\mathrm{\mu V}$)~\cite{Biondi1959AlGap} and the recently reported gap of 4Hb-TaS$_2$ ($\sim 390~\mathrm{\mu V}$)~\cite{wang2024evidence}.

\subsection{Time-domain coherence}

\begin{figure}
    \centering
    \includegraphics[width=1\linewidth]{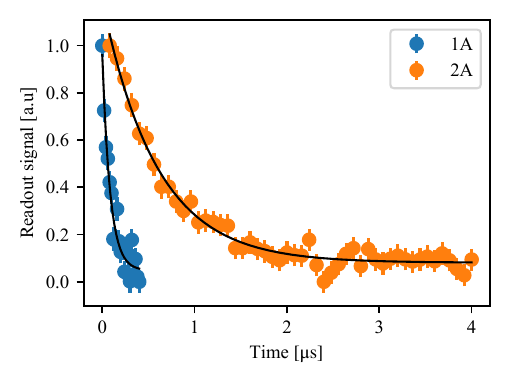}
    \caption{
    Energy relaxation of devices 1A ($T_1 = 0.08 \pm 0.01~\mu$s) and 2A ($T_1 = 0.69 \pm 0.03~\mu$s). 
    }
    \label{fig:coherence}
\end{figure}

We performed time-domain measurements on devices 1A and 2A.
Energy relaxation was measured by applying a $\pi$-pulse followed by a variable delay and readout, yielding $T_1 = 0.08 \pm 0.01~\mu$s for device 1A and $T_1 = 0.69 \pm 0.03~\mu$s for device 2A (Fig.~\ref{fig:coherence}).
Ramsey interferometry did not yield resolvable fringes. This indicates dephasing on a timescale shorter than our experimental time resolution of $16~\mathrm{ns}$, preventing a quantitative extraction of $T_2^*$ with the present pulse sequence.

\section{Discussion}

The primary outcome of this work is the demonstration that an Al/AlO$_x$/4Hb-TaS$_2$ junction fabricated by a full-oxidation layer process functions as a coherent Josephson element in a transmon, exhibiting a flux-tunable spectrum consistent with standard transmon physics and enabling time-domain control.

The Ambegaokar--Baratoff relation typically provides a reliable proxy for the Josephson energy in standard devices~[12, 13]; interestingly, here there is a factor of $\sim 5$ discrepancy.
Possible explanations for this discrepancy, arising from the unconventional superconductivity of 4Hb-TaS$_2$, include multigap structure in 4Hb-TaS$_2$
~\cite{wang2024evidence}, anisotropic or sign-changing pairing that suppresses Josephson coupling~\cite{TanakaKashiwaya1995}, interface-dependent tunneling matrix elements that selectively couple to particular bands~\cite{Brinkman2002}, and coupling to low-energy surfaces or edge modes in the 4Hb-TaS$_2$~\cite{Nagato1996, Sato2017Review}.

The observed relaxation times are below those of optimized Al/AlO$_x$/Al transmons; however, this first-generation hybrid platform is primarily motivated by access to unconventional superconductivity rather than immediate lifetime benchmarking.
In this context, it is important to identify plausible loss channels that are specific to the 4Hb-TaS$_2$ integration and that can guide future device optimization. Whether the high dephasing rate is a result of flux noise in the SQUID loops or is intrinsic to the 4Hb-TaS$_2$ requires further work. For example, an intrinsic mechanism might be quasiparticle-induced dephasing associated with enhanced subgap density of states in 4Hb-TaS$_2$, which could be exacerbated by a low effective gap energy.


A central motivation for this hybrid architecture is coupling to boundary and subgap degrees of freedom in 4Hb-TaS$_2$, including features reported in scanning tunneling spectroscopy at step edges and in vortex cores~\cite{Nayak2021Evidence}.
In the present geometry, several factors may suppress their visibility in transmon spectroscopy: supercurrent pathways dominated by bulk regions that shunt edge contributions; relevant mode energies outside the accessible measurement bandwidth; and junction separations and geometries that reduce coherent bound-state formation~\cite{Beenakker1991Universal}.
These considerations motivate future edge-sensitive junction layouts and higher-transparency weak links.
Even without resolving such modes here, establishing coherent circuit operation is a prerequisite for applying the cQED toolbox as a spectroscopic probe of this material class.
Finally, the many transitions observed at high drive power are well captured by the standard transmon–cavity model, underscoring that future searches for additional (“hidden”) transitions must account for the full coupled transmon–cavity Hamiltonian.

\section{Conclusion}

We demonstrated a flux-tunable 3D transmon qubit whose Josephson element is based on an Al/AlO$_x$/4Hb-TaS$_2$ junction fabricated via complete oxidation of a thin Al layer deposited onto an exfoliated 4Hb-TaS$_2$ flake and capped by Al.
The device exhibits a well-behaved transmon spectrum with SQUID-like tunability and sub-microsecond energy relaxation ($T_1$ in the $0.08$--$0.69~\mu\mathrm{s}$ range across measured devices), while $T_2^{*}$ could not be extracted from Ramsey measurements, as coherence decays on a timescale shorter than our $16$ ns pulse-time resolution. This rapid dephasing may be related to quasiparticle dynamics in 4Hb-TaS$_2$: a comparatively low effective gap energy (e.g., from a small-gap band or enhanced subgap density of states) could support a higher nonequilibrium quasiparticle population and thereby increase quasiparticle-induced dephasing. Further work is required to determine whether $T_2^*$ is limited intrinsically by the material or by extrinsic fabrication quality or flux noise. These results establish a practical route to integrating exfoliable vdW superconductors into coherent superconducting circuits, demonstrated here with 4Hb-TaS$_2$, and define a baseline for future circuit-QED studies that target boundary and subgap physics in this unconventional-superconductor platform.

For future work, we suggest etching the 4Hb-TaS$_2$ flake into a narrow strip for better control of the junction size and more defined edge mode coupling. Perhaps in this way a clear ABS might emerge. In addition, more work could be done to study the relation between the resistance and the Josephson energy of Al/AlO$_x$/4Hb-TaS$_2$ junctions.

\section{Acknowledgments}
We thank S. Frolov and V. Fatemi for helpful discussions. This research project was fully supported by the Helen Diller Quantum Center at the Technion.

\section{Conflict of Interest}
The authors declare no competing interests.

\section{Data availability}
All data supporting the findings of this study are available upon reasonable request.

\bibliography{bib}

@article{Dentelski2021Robust,
  title = {Robust gapless superconductivity in $4Hb\text{\ensuremath{-}}{\mathrm{TaS}}_{2}$},
  author = {Dentelski, David and Day-Roberts, Ezra and Birol, Turan and Fernandes, Rafael M. and Ruhman, Jonathan},
  journal = {Phys. Rev. B},
  volume = {103},
  issue = {22},
  pages = {224522},
  numpages = {13},
  year = {2021},
  month = {Jun},
  publisher = {American Physical Society},
  doi = {10.1103/PhysRevB.103.224522},
  url = {https://link.aps.org/doi/10.1103/PhysRevB.103.224522}
}

@article{Koch2007Transmon,
  title   = {Charge-insensitive qubit design derived from the Cooper pair box},
  author  = {Koch, Jens and Yu, Terri M. and Gambetta, Jay and Houck, A. A. and Schuster, D. I. and Majer, J. and Blais, Alexandre and Devoret, M. H. and Girvin, S. M. and Schoelkopf, R. J.},
  journal = {Phys. Rev. A},
  volume  = {76},
  number  = {4},
  pages   = {042319},
  year    = {2007},
  month   = {10},
  doi     = {10.1103/PhysRevA.76.042319},
  url     = {https://link.aps.org/doi/10.1103/PhysRevA.76.042319},
  publisher = {American Physical Society}
}

@article{Ribak2020Chiral4HbTaS2,
  title   = {Chiral superconductivity in the alternate stacking compound 4Hb-TaS\textsubscript{2}},
  author  = {Ribak, A. and Lahoud, E. and Silber, I. and Almoalem, A. and Nagai, Y. and Nakamura, Y. and Ribak, E. N. and Kanigel, A. and Beidenkopf, H.},
  journal = {Science Advances},
  volume  = {6},
  number  = {10},
  pages   = {eaax9480},
  year    = {2020},
  doi     = {10.1126/sciadv.aax9480},
  url     = {https://doi.org/10.1126/sciadv.aax9480},
  publisher = {American Association for the Advancement of Science}
}

@Article{Nayak2021Evidence,
author={Nayak, Abhay Kumar
and Steinbok, Aviram
and Roet, Yotam
and Koo, Jahyun
and Margalit, Gilad
and Feldman, Irena
and Almoalem, Avior
and Kanigel, Amit
and Fiete, Gregory A.
and Yan, Binghai
and Oreg, Yuval
and Avraham, Nurit
and Beidenkopf, Haim},
title={Evidence of topological boundary modes with topological nodal-point superconductivity},
journal={Nature Physics},
year={2021},
month={Dec},
day={01},
volume={17},
number={12},
pages={1413-1419},
issn={1745-2481},
doi={10.1038/s41567-021-01376-z},
url={https://doi.org/10.1038/s41567-021-01376-z}
}

@article{Beenakker1991Universal,
  title = {Universal limit of critical-current fluctuations in mesoscopic Josephson junctions},
  author = {Beenakker, C. W. J.},
  journal = {Phys. Rev. Lett.},
  volume = {67},
  issue = {27},
  pages = {3836--3839},
  numpages = {0},
  year = {1991},
  month = {Dec},
  publisher = {American Physical Society},
  doi = {10.1103/PhysRevLett.67.3836},
  url = {https://link.aps.org/doi/10.1103/PhysRevLett.67.3836}
}

@article{AmbegaokarBaratoff1964,
  author       = {V. Ambegaokar and A. Baratoff},
  title        = {Temperature Dependence of the Josephson Current in Superconducting Tunneling},
  journal      = {Physical Review},
  volume       = {135},
  number       = {3A},
  pages        = {A1069--A1074},
  year         = {1964},
  doi          = {10.1103/PhysRev.135.A1069},
}

@article{Almoalem2024_pi_shifts,
  author       = {Avior Almoalem and Irena Feldman and Ilay Mangel and Michael Shlafman and Yuval E. Yaish and Mark H. Fischer and Michael Moshe and Jonathan Ruhman and Amit Kanigel},
  title        = {The observation of {$\pi$}-shifts in the Little–Parks effect in 4Hb-TaS$_2$},
  journal      = {Nature Communications},
  volume       = {15},
  pages        = {4623},
  year         = {2024},
  doi          = {10.1038/s41467-024-48260-x},
  url          = {https://www.nature.com/articles/s41467-024-48260-x},
}

@misc{chistolini2025contactlesscavitysensingsuperfluid,
      title={Contactless cavity sensing of superfluid stiffness in atomically thin 4Hb-TaS$_2$}, 
      author={Trevor Chistolini and Ha-Leem Kim and Qiyu Wang and Su-Di Chen and Luke Pritchard Cairns and Ryan Patrick Day and Collin Sanborn and Hyunseong Kim and Zahra Pedramrazi and Ruishi Qi and Takashi Taniguchi and Kenji Watanabe and James G. Analytis and David I. Santiago and Irfan Siddiqi and Feng Wang},
      year={2025},
      eprint={2510.25124},
      archivePrefix={arXiv},
      primaryClass={cond-mat.supr-con},
      url={https://arxiv.org/abs/2510.25124}, 
}

@misc{wang2024evidence,
      title={Evidence for multiband gapless superconductivity in the topological superconductor candidate 4Hb-TaS2}, 
      author={Hanru Wang and Yihan Jiao and Fanyu Meng and Xu Zhang and Dongzhe Dai and Chengpeng Tu and Chengcheng Zhao and Lu Xin and Sicheng Huang and Hechang Lei and Shiyan Li},
      year={2024},
      eprint={2412.08450},
      archivePrefix={arXiv},
      primaryClass={cond-mat.supr-con},
      url={https://arxiv.org/abs/2412.08450}, 
}

@article{TanakaKashiwaya1995,
  author  = {Y. Tanaka and S. Kashiwaya},
  title   = {Theory of Josephson Effects in Anisotropic Superconductors},
  journal = {Physical Review B},
  volume  = {53},
  pages   = {9371--9380},
  year    = {1996},
  doi     = {10.1103/PhysRevB.53.9371}
}

@article{Brinkman2002,
  author  = {A. Brinkman and A. A. Golubov and M. Y. Kupriyanov},
  title   = {Multiband model for tunneling in MgB2 junctions},
  journal = {Physical Review B},
  volume  = {65},
  pages   = {180517},
  year    = {2002},
  doi     = {10.1103/PhysRevB.65.180517}
}

@article{Nagato1996,
  author  = {Y. Nagato and K. Nagai and J. Hara},
  title   = {Theory of Andreev Bound States and Tunneling Spectroscopy in Unconventional Superconductors},
  journal = {Journal of Low Temperature Physics},
  volume  = {103},
  pages   = {1--24},
  year    = {1996},
  doi     = {10.1007/BF00754048}
}

@article{Sato2017Review,
  author  = {Masatoshi Sato and Yukio Ando},
  title   = {Topological superconductors: a review},
  journal = {Reports on Progress in Physics},
  volume  = {80},
  pages   = {076501},
  year    = {2017},
  doi     = {10.1088/1361-6633/aa6ac7}
}

@article{Biondi1959AlGap,
  author  = {Biondi, Manfred A. and Garfunkel, M. P.},
  title   = {Millimeter Wave Absorption in Superconducting Aluminum. I. Temperature Dependence of the Energy Gap},
  journal = {Physical Review},
  volume  = {116},
  number  = {4},
  pages   = {853--861},
  year    = {1959},
  month   = nov,
  doi     = {10.1103/PhysRev.116.853},
  url     = {https://link.aps.org/doi/10.1103/PhysRev.116.853},
  note    = {Determines the superconducting gap in Al; reports $\Delta(0)=(3.2\pm0.1)\,k_B T_c$ with $T_c=1.178$ K.}
}

@article{Antony2021MiniaturizingTransmon,
  title        = {Miniaturizing {T}ransmon Qubits Using van der {W}aals Materials},
  author       = {Antony, Abhinandan and Gustafsson, Martin V. and Ribeill, Guilhem J. and Ware, Matthew and Rajendran, Anjaly and Govia, Luke C. G. and Ohki, Thomas A. and Taniguchi, Takashi and Watanabe, Kenji and Hone, James and Fong, Kin Chung},
  journal      = {Nano Letters},
  year         = {2021},
  volume       = {21},
  number       = {23},
  pages        = {10122--10126},
  month        = nov,
  doi          = {10.1021/acs.nanolett.1c04160},
  url          = {https://doi.org/10.1021/acs.nanolett.1c04160},
  publisher    = {American Chemical Society}
}

@misc{Balgley2025CoherentCompactVdWTransmons,
  title         = {Coherent and compact van der Waals transmon qubits},
  author        = {Balgley, Jesse and Park, Jinho and Chu, Xuanjing and Liu, Jiru and Holbrook, Madisen and Watanabe, Kenji and Taniguchi, Takashi and Kamal, Archana and Ranzani, Leonardo and Gustafsson, Martin V. and Hone, James and Fong, Kin Chung},
  year          = {2025},
  eprint        = {2512.08059},
  archivePrefix = {arXiv},
  primaryClass  = {quant-ph},
  doi           = {10.48550/arXiv.2512.08059}
}

\end{document}